\documentclass[12pt]{article}

\usepackage{amsmath,amssymb}
\usepackage{bm}
\usepackage{graphicx, color}
\usepackage{wrapfig}
\usepackage{cite}
\begin{document}
\title{
\begin{flushright}
\ \\*[-80pt] 
\begin{minipage}{0.2\linewidth}
\normalsize
%arXiv:YYMM.NNNN \\
HUPD1908 \\*[50pt]
\end{minipage}
\end{flushright}
{\Large \bf 
$A_4$ flavor symmetric model in SUSY SU(5) GUT
\\*[20pt]}}

\author{ 
\centerline{
Yu~Muramatsu $^{1}$\footnote{yumura@mail.ccnu.edu.cn},
~Hideaki~Okane $^{2}$\footnote{hideaki-ookane@hiroshima-u.ac.jp},
~Yusuke~Shimizu $^{2}$\footnote{yu-shimizu@hiroshima-u.ac.jp},
~Kenta~Takagi $^{2}$\footnote{takagi-kenta@hiroshima-u.ac.jp}}
\\*[20pt]
\centerline{
\begin{minipage}{\linewidth}
\begin{center}
$^1${\it \normalsize
Central China Normal University, Wuhan, Hubei 430079, People's Republic of China} \\*[5pt]
$^2${\it \normalsize
Graduate School of Science, Hiroshima University, Higashi-Hiroshima 739-8526, Japan}
\end{center}
\end{minipage}}
\\*[50pt]}

\date{
\centerline{\small \bf Abstract}
\begin{minipage}{0.9\linewidth}
\medskip 
\medskip 
\small 
We propose a model with $A_4$ flavor symmetry for leptons and quarks in the framework of supersymmetric SU(5) grand unified theory (GUT). The running masses of quarks and charged leptons at GUT scale ($\sim 10^{16}$ GeV) are realized by the adjoint 24-dimensional Higgs multiplet and additional gauge singlet scalar fields including flavons. In this paper, we focus on a result of the quark and charged lepton masses and quark mixing since our present model is known to reproduce recent experimental results of the neutrino mass and oscillation. Those results are showed numerically.
\end{minipage}
}

\begin{titlepage}
\maketitle
\thispagestyle{empty}
\end{titlepage}
\newpage
%------------------------------------------------------------------------------%
%--------------------------------   Introduction   --------------------------------%
%------------------------------------------------------------------------------%
\section{Introduction}
The success of the standard model (SM) is widely confirmed by various experiments.
However, there are some mysterious origins which can not be explained by the SM.
The origins of flavors of fermions and charges of SM particles are famous examples.
The remarkable developments in the neutrino oscillation experiments have fueled us to make phenomenological models which realize the lepton flavor mixing in various ways.
It is popular method to assume some flavor symmetry of $S_3$, $A_4$, $S_4$, $A_5$ and other large non-Abelian discrete groups~\cite{Altarelli:2010gt,Ishimori:2010au,Ishimori:2012zz,King:2013eh,King:2014nza}.
Since it is difficult to determine the true flavor model from within or out of them, 
it is important to test them by different observations not only by the neutrino oscillation.
Then, we extend a lepton flavor model to explain quark mixing with the SU(5) framework and compare its results and observed Cabibbo-Kobayashi-Maskawa (CKM) parameters~\cite{Charles:2004jd} and running masses of quarks and leptons~\cite{Ross:2007az}.
In the GUT, the unification group is an origin of charges of SM particles~\cite{Georgi:1974sy}.
However, it is a problem that the charged leptons and down-type quarks are unified in SU(5) representation,
which leads to the wrong prediction of the charged lepton and down-type quark masses~\cite{Buras:1977yy}.
To consider contributions from a Higgs multiplet of the adjoint representation is a popular method to make a difference among the charged lepton masses and down-type quark masses in order to avoid this problem~\cite{Ellis:1979fg}.

In this paper, we consider a model with $A_4$ flavor symmetry because it is a minimal group which includes triplet irreducible representation and can explain three flavors of fermions naturally.
In addition, our model has $Z_3$, $U(1)_{FN}$ and $U(1)_R$ symmetries which are necessary to forbid unwanted couplings and explain the hierarchical masses of fermions.
We note that there is no $U(1)_{FN}$ charge in the neutrino sector and we already know that the results for neutrino mixing parameters of this model~\cite{Shimizu:2011xg} are consistent with the current experimental data~\cite{Esteban:2016qun}.
Therefore, we focus only on the results of quark mixing parameters and running masses of quarks and leptons.

As we said above, the GUT explains the origin of charges of SM particles, on the other hand it is not easy to realize observed fermion masses and mixing angles.
Therefore, it is interesting to explain the flavor and charge puzzle of fermions by a flavor symmetry in a specific GUT framework \cite{Antusch:2011qg, King:2017guk}.
An earlier attempt to extend an $A_4$ flavor model in the supersymmetric (SUSY) SU(5) framework~\cite{Altarelli:2008bg} proposed a simple model to realize required quark mixing and tri-bimaximal mixing (TBM) for neutrinos.
The reactor experiments measured non-zero value of $\theta_{13}$ in 2012~\cite{An:2012eh,Ahn:2012nd} and excluded TBM pattern of neutrino mixing.
The non-zero value of $\theta_{13}$ was realized by an $A_4$ symmetry in the SU(5) GUT in Refs.~\cite{Bjorkeroth:2015ora, Belyaev:2018vkl, Laamara:2018zpo, Bernigaud:2018qky}.

In our present model, we use the 24-dimensional adjoint Higgs multiplet $H_{24}$ contribution to make a difference between masses of down-quarks and charged leptons.
Moreover, we introduce a Froggatt-Nielsen (FN) field~\cite{Froggatt:1978nt}, $\Theta$, to realize the observed fermion mass hierarchy.
The flavons, $\phi_T$, $\phi_S$, $\xi$ and $\xi'$, are also introduced in order to realize the latest neutrino oscillation experiments.
It is noted that $\xi$ and $\xi'$ are coupled to the quarks not only to the leptons.
If we determine the vacuum expectation values (VEVs) of $\xi$ and $\xi'$ from quark sector,
we cannot find any contribution to the lepton mixing parameters because these VEVs are absorbed in the corresponding Yukawa coupling constants.
The VEVs of those scalar fields determine the masses of quarks and leptons for different flavors and explain the flavor mixing.
The required VEV alignments are derived to minimize the scalar potential, which is shown in this paper.
We also show the numerical result of quark mixing and masses of quarks and charged leptons at GUT scale ($\sim10^{16}$ GeV).

The rest of this paper is composed as follows.
In section 2, we give our framework of $A_4$ flavor model in SUSY SU(5) GUT.
We also provide the potential analysis for corresponding scalar fields in this section.
In section 3, our numerical results for the quark and lepton masses and the quark mixing is shown.
Summary and discussions are given in section 4.
We give a brief explanation of the multiplication rule of $A_4$ in Appendix A.
In Appendix B, we show some relations among the fermion mass ratios and the model parameters used in the numerical analysis.

%------------------------------------------------------------------------------%
%------------------------------------------------------------------------------%
%------------------------------------------------------------------------------%
\section{A framework of SUSY SU(5) GUT}
We illustrate our model in this section.
In order to explain the flavor structure of quarks and leptons simultaneously,
we consider a $A_4$ flavor model in SUSY SU(5) GUT.
In addition, we also impose $Z_3$, $U(1)_{FN}$ and $U(1)_R$ symmetries on our model to eliminate unwanted couplings.
%------------------------------------------------------------------------------%
%------------------------------------------------------------------------------%
\subsection{Particle contents and mass terms}
We introduce some SU(5) singlet scalar fields such as $\phi_T$, $\phi_S$, $\xi$, $\xi'$ and $\Theta$ to realize the observed fermion mass hierarchy and flavor mixing.
The fermion masses are proportional to the VEVs of these scalar fields.
The driving fields $\phi_0^T$, $\phi_0^S$ and $\xi_0$ are necessary to get non-zero VEVs of the SU(5) singlet scalar fields.
The VEVs of driving fields are zero.
The charge assignment of all the fermions and relevant scalar fields are summarized in Table \ref{tb:fields}.
We also assume Majorana neutrinos and the type-I seesaw mechanism~\cite{Minkowski:1977sc, Yanagida, GellMann:1980vs, Mohapatra:1979ia, Schechter:1980gr} with additional three heavy right-handed neutrinos.
%The quarks and leptons are described as the SU(5) representation:
%\begin{align}
%	F({\bf \bar{5}})=\begin{pmatrix}d_{1R}^c\\d_{2R}^c\\d_{3R}^c\\e\\-\nu\end{pmatrix}_L,\quad
%	T({\bf 10})=\begin{pmatrix}
%	0&u_{3R}^c&-u_{2R}^c&u_{1L}&d_{1L} \\
%	-u_{3R}^c&0&u_{1R}^c&u_{2L}&d_{2L} \\
%	u_{2R}^c&-u_{1R}^c&0&u_{3L}&d_{3L} \\
%	-u_{1L}&-u_{2L}&-u_{3L}&0&e_R^c \\
%	-d_{1L}&-d_{2L}&-d_{3L}&-e_R^c&0
%	\end{pmatrix}_L,\quad
%	N({\bf 1})=\nu_R^c,
%\end{align}
%where we omit the flavor indices.
%------------------    Table   --------------------%
\begin{table}[b]
\centering
 	\begin{tabular}{|c||c|c|c|c|c||c|c|c|c|c|c|c|c||c|c|c|} \hline 
		\rule[14pt]{0pt}{0pt}
		&$T_1$&$T_2$&$T_3$&$F$&$N$&$H_5$&$H_{\bar{5}}$&$H_{24}$&$\phi_T$&$\phi_S$&$\xi$&$\xi' $&$\Theta$&$\phi_0^T$&$\phi_S^0$&$\xi_0$\\ \hline \hline 
		\rule[14pt]{0pt}{0pt}
		$SU(5)$&$10$&$10$&$10$&$\bar{5}$&$1$&$5$&$\bar{5}$&$24$&$1$&$1$&$1$&$1$&1&1&1&1\\

		$A_4$&$1$&$1''$&$1'$&$3$&$3$&$1$&$1$&$1$&$3$&$3$&$1$&$1'$&1&3&3&1\\

		$Z_3$&$\omega^2$&$\omega^2$&$\omega^2$&$\omega$&$\omega^2$&1&1&1&1&$\omega^2$&$\omega^2$&$\omega^2$&1&1&$\omega^2$&$\omega^2$\\

		$U(1)_{FN}$&3&2&0&0&0&0&0&0&0&0&0&0&-1&0&0&0\\
		
		$U(1)_{R}$&1&1&1&1&1&0&0&0&0&0&0&0&0&2&2&2\\ \hline
		
	\end{tabular}
\caption[]{\parbox[t]{0.85\hsize}
{The charge assignment of $SU(5)$, $A_4$, $Z_3$, $U(1)_{FN}$ and $U(1)_R$ for all the relevant particles of our model ($\omega=e^{2\pi i/3}$).}}
\label{tb:fields}
\end{table}
%---------------------------------------------------%
Based on this set-up, we obtain the superpotential for Yukawa interactions,
\begin{align}
	w=w_{\bar{5}}+w_{10}+w_D+w_N,
\end{align}
where $w_{\bar{5}}$, $w_{10}$, $w_D$ and $w_N$ stand for the mass terms of down-type fermions, up-type quarks, Dirac- and Majorana-type of neutrinos respectively.

At first, $w_D$ and $w_N$ are written as follows:
\begin{align}
	w_D=&y^DFNH_5, \\
	w_N=&y^N_1NN\phi_S+y^N_2NN\xi+y^N_3NN\xi'.
\end{align}
We note that they are same as the neutrino mass terms shown in the previous work~\cite{Morozumi:2017rrg} whose predictions are consistent with the current neutrino oscillation experiments~\cite{Esteban:2016qun} within $3\sigma$ error.

The mass terms of charged leptons and down-type quarks are written in a unified form as
\begin{align}
	w_{\bar{5}}=(y^{\bar{5}}_3+y^{24}_3\frac{H_{24}}{\Lambda})FT_3H_{\bar{5}}\frac{\phi_T}{\Lambda}+
	(y^{\bar{5}}_2+y^{24}_2\frac{H_{24}}{\Lambda})FT_2H_{\bar{5}}\frac{\phi_T}{\Lambda}\frac{\Theta^2}{\Lambda^2}+
	(y^{\bar{5}}_1+y^{24}_1\frac{H_{24}}{\Lambda})FT_1H_{\bar{5}}\frac{\phi_T}{\Lambda}\frac{\Theta^3}{\Lambda^3}.
\label{eq:w5}
\end{align}
It leads to a diagonal mass matrix if the VEV of $\phi_T$ is same as the previous works' in accordance with the multiplication rule of $A_4$ shown in Appendix~A.
The VEV of 24-dimensional adjoint Higgs $H_{24}$ realizes different mass eigenvalues for the charged leptons and down-type quarks.
The observed mass hierarchy is realized by FN field $\Theta$ coupled in different orders for each term.

We also note that the previous work~\cite{Morozumi:2017rrg} require the charged lepton mass matrix to be diagonal.
Moreover, we can obtain the same VEV alignments for $\phi_S$ and $\phi_T$ as the previous work, which is shown in the next section.
Therefore, we predict the same lepton mixing matrix, Pontecorvo-Maki-Nakagawa-Sakata(PMNS) matrix, as the previous works' and we do not discuss the lepton mixing in the following.

The mass term of up-type quarks is obtained as follows:
\begin{align}
\begin{aligned}
	w_{10}=&(y^5_{11}+y^{24}_{11}\frac{H_{24}}{\Lambda})T_1T_1H_5\frac{\xi}{\Lambda}\frac{\Theta^6}{\Lambda^6}+
	(y^5_{22}+y^{24}_{22}\frac{H_{24}}{\Lambda})T_2T_2H_5\frac{\xi}{\Lambda}\frac{\xi'^2}{\Lambda^2}\frac{\Theta^4}{\Lambda^4}\\
	&+(y^5_{33}+y^{24}_{33}\frac{H_{24}}{\Lambda})T_3T_3H_5\frac{\xi'}{\Lambda}+
	(y^5_{12}+y^{24}_{12}\frac{H_{24}}{\Lambda})T_1T_2H_5\frac{\xi'}{\Lambda}\frac{\Theta^5}{\Lambda^5}\\
	&+(y^5_{13}+y^{24}_{13}\frac{H_{24}}{\Lambda})T_1T_3H_5\frac{\xi}{\Lambda}\frac{\xi'^2}{\Lambda^2}\frac{\Theta^3}{\Lambda^3}
	+(y^5_{23}+y^{24}_{23}\frac{H_{24}}{\Lambda})T_2T_3H_5\frac{\xi}{\Lambda}\frac{\Theta^2}{\Lambda^2}.
\end{aligned}
\end{align}
%The terms with $H_{24}$ make no change in mass matrix structure but only overall factors
%under the following vacuum expectation values.
%\begin{align}
%	\langle H_5\rangle=\langle\begin{pmatrix}T_1\\T_2\\T_3\\H^+\\H_0\end{pmatrix}\rangle,
%	\langle H_{\bar{5}}\rangle=\langle\begin{pmatrix}T_1^c\\T_2^c\\T_3^c\\H^-\\-H_0\end{pmatrix}\rangle,
%	\langle H_{24}\rangle=\frac{2}{\sqrt{30}}v_{24}^H\begin{pmatrix}1&&&&\\&1&&&\\&&1&&\\&&&-3/2&\\&&&&-3/2\end{pmatrix}.
%\end{align}
We can reduce it by introducing new parameters, such as $y_{11}^u$, $y_{12}^u$,\dots, to the following form:
\begin{align}
\begin{aligned}
	w_{10}=&y^u_{11}T_1T_1H_5\frac{\xi}{\Lambda}\frac{\Theta^6}{\Lambda^6}
	+y^u_{22}T_2T_2H_5\frac{\xi}{\Lambda}\frac{\xi'^2}{\Lambda^2}\frac{\Theta^4}{\Lambda^4}
	+y^u_{33}T_3T_3H_5\frac{\xi'}{\Lambda} \\
	&+y^u_{12}T_1T_2H_5\frac{\xi'}{\Lambda}\frac{\Theta^5}{\Lambda^5}+
	y^u_{13}T_1T_3H_5\frac{\xi}{\Lambda}\frac{\xi'^2}{\Lambda^2}\frac{\Theta^3}{\Lambda^3}+
	y^u_{23}T_2T_3H_5\frac{\xi}{\Lambda}\frac{\Theta^2}{\Lambda^2}.
\label{eq:w10}
\end{aligned}
\end{align}
In addition to $\Theta$, the VEVs of $\xi$ and $\xi '$ contribute the quark mixing matrix.
We note that the VEVs of $\xi$ and $\xi '$ determined in the quark sector do not affect on the lepton mixing because of the freedom of reparametrization of Yukawa couplings in the neutrino sector.
Therefore, we can discuss the quark mixing and masses independently from the neutrino sector whose results are consistent to the current experimental data of neutrino oscillations.
%------------------------------------------------------------------------------%
%------------------------------------------------------------------------------%
\subsection{Potential analysis}
In order to obtain finite fermion masses, the scalar fields coupled to fermions must have non-zero VEVs.
In this section, we show a derivation of required VEVs with the following potential analysis.
We denote the superpotential for the scalar fields of our present model as $w_d$ and write it in two parts:
\begin{align}
	w_d\equiv w_d^T+w_d^S,
%	+w_d^\Theta
	\label{eq:wd}
\end{align}
where
\begin{align}
\begin{aligned}
	&w_d^T=-M(\phi_0^T\phi_T)_{\bf 1}+g\phi_0^T(\phi_T\phi_T)_{\bf 3}, \\
	&w_d^S=
	g_1\phi_0^S(\phi_S\phi_S)_{\bf 3}+
	g_2(\phi_0^S\phi_S)_{\bf 1}\xi+g_2'(\phi_0^S\phi_S)_{\bf 1''}\xi '+
	g_3(\phi_S\phi_S)_{\bf 1}\xi_0-
	g_4\xi_0\xi\xi.
%	\\&w_d^\Theta=h(\Theta_0\phi_S)_{\bf 1}\Theta-h'(\Theta_0\phi_S)_{\bf 1''}\Theta'.
\end{aligned}
\end{align}
The scalar fields appeared above obtain their VEVs at the point where the scalar potential is minimum.
We denote the scalar potential as $V$ and separate it into two parts: $V=V_F+V_D$, where $V_F$ and $V_D$ are the scalar potential of $F$- and $D$- terms respectively.
The $F$- term, $V_F=V_T+V_S$, is given by
\begin{align}
\begin{aligned}
	V_T=\sum_{X}\left|\frac{\partial w_d^T}{\partial X}\right|^2
	=&\left| -M\phi_{T1}+\frac{2}{3}g(\phi_{T1}^2-\phi_{T2}\phi_{T3})\right|^2 \\
	+&\left| -M\phi_{T3}+\frac{2}{3}g(\phi_{T2}^2-\phi_{T3}\phi_{T1})\right|^2 \\
	+&\left| -M\phi_{T2}+\frac{2}{3}g(\phi_{T3}^2-\phi_{T1}\phi_{T2})\right|^2 \\
	+&\left| -M\phi_{01}^T+\frac{2}{3}g(2\phi_{01}^T\phi_{T1}-\phi_{03}^T\phi_{T2}-\phi_{02}^T\phi_{T3})\right|^2 \\
	+&\left| -M\phi_{03}^T+\frac{2}{3}g(2\phi_{02}^T\phi_{T2}-\phi_{01}^T\phi_{T3}-\phi_{03}^T\phi_{T1})\right|^2 \\
	+&\left| -M\phi_{02}^T+\frac{2}{3}g(2\phi_{03}^T\phi_{T3}-\phi_{02}^T\phi_{T1}-\phi_{01}^T\phi_{T2})\right|^2 ,
\end{aligned}\label{eq:VT}
\end{align}
and
\begin{align}
\begin{aligned}
	V_S=\sum_{Y}\left|\frac{\partial w_d^S}{\partial Y}%(w_d^S+w_d^\Theta)
	\right|^2
	=&\left|\frac{2}{3}g_1(\phi_{S1}^2-\phi_{S2}\phi_{S3})-g_2\phi_{S1}\xi+g_2'\phi_{S3}\xi'\right|^2 \\
	+&\left|\frac{2}{3}g_1(\phi_{S2}^2-\phi_{S3}\phi_{S1})-g_2\phi_{S3}\xi+g_2'\phi_{S2}\xi'\right|^2 \\
	+&\left|\frac{2}{3}g_1(\phi_{S3}^2-\phi_{S1}\phi_{S2})-g_2\phi_{S2}\xi+g_2'\phi_{S1}\xi'\right|^2 \\
	+&\left|g_3(\phi_{s1}^2+2\phi_{S2}\phi_{S3})-g_4\xi^2\right|^2 \\
	%
%	+&\left|h\Theta\phi_{S1}-h'\Theta'\phi_{S3}\right|^2
%	+\left|h\Theta\phi_{S3}-h'\Theta'\phi_{S2}\right|^2
%	+\left|h\Theta\phi_{S2}-h'\Theta'\phi_{S1}\right|^2 \\
	%
	+&\left|\frac{2}{3}g_1(2\phi_{01}^S\phi_{S1}-\phi_{03}^S\phi_{S2}-\phi_{02}^S\phi_{S3})
	-g_2\phi_{01}^S\xi+g_2'\phi_{03}^S\xi'+2g_3\phi_{S1}\xi_0
%	+h\Theta\Theta_{01}-h'\Theta'\Theta_{03}
	\right|^2 \\
	+&\left|\frac{2}{3}g_1(2\phi_{02}^S\phi_{S2}-\phi_{01}^S\phi_{S3}-\phi_{03}^S\phi_{S1})
	-g_2\phi_{03}^S\xi+g_2'\phi_{02}^S\xi'+2g_3\phi_{S3}\xi_0
%	+h\Theta\Theta_{03}-h'\Theta'\Theta_{02}
	\right|^2 \\
	+&\left|\frac{2}{3}g_1(2\phi_{03}^S\phi_{S3}-\phi_{02}^S\phi_{S1}-\phi_{01}^S\phi_{S2})
	-g_2\phi_{02}^S\xi+g_2'\phi_{01}^S\xi'+2g_3\phi_{S2}\xi_0
%	+h\Theta\Theta_{02}-h'\Theta'\Theta_{01}
	\right|^2 \\
	+&\left|-g_2(\phi_{01}^S\phi_{S1}+\phi_{03}^S\phi_{S2}+\phi_{02}^S\phi_{S3})-2g_4\xi\xi_0\right|^2 \\
	+&\left|g_2'(\phi_{02}^S\phi_{S2}+\phi_{01}^S\phi_{S3}+\phi_{03}^S\phi_{S1})\right|^2.
%	+&\left|h(\Theta_{01}\phi_{S1}+\Theta_{03}\phi_{S2}+\Theta_{02}\phi_{S3})\right|^2
%	+\left|h'(\Theta_{03}\phi_{S1}+\Theta_{02}\phi_{S2}+\Theta_{01}\phi_{S3})\right|^2.
\end{aligned}\label{eq:VS}
\end{align}
The sum for $X,Y$ runs over all the scalar fields of our present model:
\begin{align}
\begin{aligned}
	X=&\{\phi_{01}^T, \phi_{02}^T, \phi_{03}^T, \phi_{T1}, \phi_{T2}, \phi_{T3}\}, \\
	Y=&\{\phi_{01}^S, \phi_{02}^S, \phi_{03}^S, \xi_0, %\Theta_{01}, \Theta_{02}, \Theta_{03},
	\phi_{S1}, \phi_{S2}, \phi_{S3}, \xi, \xi '%, \Theta, \Theta'
	\}.
\end{aligned}
\nonumber
\end{align}
The $F$-terms $V_F$ is minimized at $V_T=V_S=0$,
which is realized by the following VEV alignments\footnote{
	There are still other solutions for $V_F=0$ where all the components of $\phi_T$ or $(\phi_S, \xi, \xi')$ vanish.
	The latter case is realized by $u=0$.
	The vanishing of $\phi_T$ or $(\phi_S, \xi, \xi')$ leads to the vanishing of
	down type fermion masses or seesaw neutrino masses respectively.
	Apart from such non-realistic solutions, there are some solutions for $V_F=0$ with non-zero VEVs of the driving fields.
	This case leads to the breakdown of $U(1)_R$ symmetry.
	In this paper, we only discuss the case where  $U(1)_R$ symmetry is conserved.}:
\begin{align}
	\frac{\langle\phi_T\rangle}{\Lambda}=v_T\begin{pmatrix}1\\0\\0\end{pmatrix},\,
	\frac{\langle\phi_S\rangle}{\Lambda}=v_S\begin{pmatrix}1\\1\\1\end{pmatrix},\,
	\frac{\langle\xi\rangle}{\Lambda}=u,\,
	\frac{\langle\xi'\rangle}{\Lambda}=\frac{g_2}{g_2'}u,\,
	\langle\phi_0^T\rangle=\langle\phi_0^S\rangle=
%	\langle\Theta_0\rangle=
	\begin{pmatrix}0\\0\\0\end{pmatrix},
\label{eq:VEV}
\end{align}
where we have defined the coefficients in front of triplet VEV alignments as
\begin{align}
v_T\equiv\frac{3M}{2g},\quad v_S\equiv\sqrt{\frac{g_4}{3g_3}}u.
\end{align}
%The VEVs of $\Theta$ and $\Theta'$ satisfy the following relation to minimize $V$:
%\begin{align}
%	\langle\Theta\rangle=\lambda,\quad
%	\langle\Theta'\rangle=\frac{h}{h'}\lambda.
%\end{align}
The VEV of $\Theta$ is generated in the minimization of $D$-terms:
\begin{align}
	V_D=\frac12(M_{FI}^2-g_{FN}|\Theta|^2)^2,\quad
	\frac{\langle\Theta\rangle}{\Lambda}=\pm\frac{M_{FI}}{\sqrt{g_{FN}}\Lambda}\equiv\lambda.
\label{eq:Dterm}
\end{align}
Finally, the VEVs of Higgs fields are
\begin{align}
    \langle H_5\rangle=\begin{pmatrix}0\\0\\0\\0\\v^H_5\end{pmatrix},\quad
    \langle H_{\bar{5}}\rangle=\begin{pmatrix}0\\0\\0\\0\\v^H_{\bar{5}}\end{pmatrix},\quad
    \langle H_{24}\rangle=v\mbox{diag}[2,2,2,-3,-3],
\label{eq:Higgs}
\end{align}
where $v^H_{5}\equiv v_u/\sqrt{2}$ and $v^H_{\bar{5}}\equiv v_d/\sqrt{2}$ satisfy $\sqrt{v_d^2+v_u^2}=174$\,GeV.
%------------------------------------------------------------------------------%
%------------------------------------------------------------------------------%
\subsection{Mass matrix}
We construct the mass matrices for quarks and charged leptons based on the superpotentials for their mass terms in Eqs.~\eqref{eq:w5} and \eqref{eq:w10} in the vacuum of Eqs.~\eqref{eq:VEV} and \eqref{eq:Dterm}-\eqref{eq:Higgs}.
We calculate them according to the multiplication rule of $A_4$ shown in Appendix A.
The mass matrix of up-type quarks is symmetric:
\begin{align}
	M_u=v^H_5\begin{pmatrix}
	y^u_{11}u \lambda^6&y^u_{12}u' \lambda^5&y^u_{13}u'^2 \lambda^3\\
	&y^u_{22}u'^2 \lambda^4&y^u_{23}u \lambda^2\\
	&&y^u_{33}u'
	\end{pmatrix}.
\end{align}
The mass matrix for down-type quarks and charged leptons is diagonal:
\begin{align}
	M_d=v^H_{\bar{5}}v_T\begin{pmatrix}
	(y^{\bar{5}}_1+\alpha y^{24}_1)\lambda^3&&\\
	&(y^{\bar{5}}_2+\alpha y^{24}_2)\lambda^2&\\
	&&(y^{\bar{5}}_3+\alpha y^{24}_3)
	\end{pmatrix},
\label{eq:Md}
\end{align}
where $y^{24}_i$ is redefined in this equation as $vy^{24}_i/\Lambda\rightarrow y^{24}_i$.
Therefore, the coefficient $\alpha$ takes different values in accordance with the VEV of $H_{24}$ in Eq.~\eqref{eq:Higgs} as follows:
\begin{align}
	\alpha=\begin{cases}
	\frac2{\sqrt{30}} & \hbox{for down-type quarks}\\
	-\frac3{\sqrt{30}} & \hbox{for charged leptons} 
	\end{cases}.
\label{eq:alpha}
\end{align}
It make the mass eigenvalues of charged leptons and down-type quarks different.
%------------------------------------------------------------------------------%
%------------------------------------------------------------------------------%
%------------------------------------------------------------------------------%
\section{Quark mixing and mass ratio}
We show how to predict the quark mixing and masses.
We start our analysis with an assumption for the sake of simplicity:
the VEVs such as $u,u'(\equiv g_2u/g'_2),\lambda$ are real and fixed to
\begin{align}
    u=u'=0.5,\quad \lambda=0.22.
\end{align}
Therefore, there are 12 free complex parameters which contribute to fermion mass ratio and mixing matrix:
Yukawa coupling constants, $y_i^{\bar 5}, y_i^{\bar 24}$ and $y^u_{ij}$ ($i,j=1,2,3$).
In our analysis, we consider the case with $\tan{\beta}\equiv v_u/v_d=50 $
for the validity of the prediction of the neutrino sectors in the present model\footnote{
	The numerical analysis of the neutrino sector refers to Ref. \cite{Morozumi:2017rrg} where the predictions are consistent with the observed data at the low energy scale.
	Although our model should be analyzed at the GUT scale, we don't discuss the quantum corrections for the neutrino sector in this paper.
}.
We comment on the proton decay through the dimension 5 operator.
It is known that the fast proton decay can be realized for large tan$\beta$ in the SUSY SU(5) GUT model~\cite{Goto:1998qg}.
However, proton decay operator $FTTT$ is prohibited by $U(1)_R$ symmetry.

We have found a parameter set which leads to a consistent prediction with experiments.
Our numerical results we show in the following are based on a parameter set listed in Table~\ref{tb:yukawa}.
%------------------------------------------------------------------------------%
\begin{table}[h]
\begin{center}
\begin{tabular}{cc}
\begin{minipage}{0.3\hsize}
\begin{center}
 	\begin{tabular}{|c||r|} \hline\rule[13pt]{0pt}{0pt}
		$y^u_{11}$&$-0.831 - i 0.417$\\ \hline\rule[13pt]{0pt}{0pt}
        $y^u_{12}$&$0.278 - i 0.615$\\ \hline\rule[13pt]{0pt}{0pt}
        $y^u_{13}$&$-0.704 + i 1.025$\\ \hline\rule[13pt]{0pt}{0pt}
        $y^u_{22}$&$0.554 + i 0.309$\\ \hline\rule[13pt]{0pt}{0pt}
        $y^u_{23}$&$-0.794 - i 0.282$\\ \hline\rule[13pt]{0pt}{0pt}
        $y^u_{33}$&$-0.775 + i 0.632$\\ \hline
	\end{tabular}
\end{center}
\end{minipage}
\begin{minipage}{0.3\hsize}
\begin{center}
 	\begin{tabular}{|c||r|} \hline\rule[13pt]{0pt}{0pt}
		$y^{\bar{5}}_{1}$&$0.050 - i 0.002$\\ \hline\rule[13pt]{0pt}{0pt}
        $y^{\bar{5}}_{2}$&$-0.298 + i 0.545$\\ \hline\rule[13pt]{0pt}{0pt}
        $y^{\bar{5}}_{3}$&$0.025 - i 0.943$\\ \hline\rule[13pt]{0pt}{0pt}
        $y^{24}_{1}$&$0.086 + i 0.042$\\ \hline\rule[13pt]{0pt}{0pt}
        $y^{24}_{2}$&$0.136 - i 0.952$\\ \hline\rule[13pt]{0pt}{0pt}
        $y^{24}_{3}$&$0.999 + i 0.039$\\ \hline
	\end{tabular}
\end{center}
\end{minipage}
\end{tabular}
\caption[]{\parbox[t]{0.85\hsize}
{A parameter set used for the numerical results shown in this paper. Yukawa parameters in the left panel contribute to the up-type quark mass matrix. The right to the down-type quarks and charged lepton.}}
\label{tb:yukawa}
\end{center}
\end{table}
%------------------------------------------------------------------------------%
Hierarchical values for down-type Yukawa coupling constants are required in order to realize the global fit data of running mass ratios among down-type quarks as well as charged leptons at GUT scale~\cite{Ross:2007az}.
On the other hand, up-type Yukawa coupling constants are allowed for non-hierarchical values.
The predicted fermion mass ratios at GUT scale are listed in Table~\ref{tb:massratio} but for neutrinos.
%------------------------------------------------------------------------------%
\begin{table}[h]
\centering
 	\begin{tabular}{|c|c||c|c||c|c|} \hline
		\rule[14pt]{0pt}{0pt}
		$m_d/m_s$&$m_s/m_b$&$m_u/m_c$&$m_c/m_t$&$m_e/m_\mu$&$m_\mu/m_\tau$\\ \hline\hline
        $0.057$&$0.0153$&$0.0033$&$0.0023$&$0.0050$&$0.0499$\\ \hline
	\end{tabular}
\caption[]{\parbox[t]{0.85\hsize}
{The predicted running mass ratio at GUT scale obtained from the parameter sets in Table~\ref{tb:yukawa}.}}
\label{tb:massratio}
\end{table}
%------------------------------------------------------------------------------%
These results are consistent with the global fit~\cite{Ross:2007az} for $\tan{\beta}=50$ case.

Next, we show the predicted CKM matrix.
The CKM matrix is parametrized by three mixing angles $\theta_{ij}$ and one CP violating phase $\delta_{CP}$ as follows:
\begin{align}
V_\text{CKM} =
\begin{pmatrix}
c_{12} c_{13} & s_{12} c_{13} & s_{13}e^{-i\delta _\text{CP}} \\
-s_{12} c_{23} - c_{12} s_{23} s_{13}e^{i\delta _\text{CP}} &
c_{12} c_{23} - s_{12} s_{23} s_{13}e^{i\delta _\text{CP}} & s_{23} c_{13} \\
s_{12} s_{23} - c_{12} c_{23} s_{13}e^{i\delta _\text{CP}} &
-c_{12} s_{23} - s_{12} c_{23} s_{13}e^{i\delta _\text{CP}} & c_{23} c_{13}
\end{pmatrix}.
\label{eq:CKM}
\end{align}
where $c_{ij}$ and $s_{ij}$ denote $\cos\theta_{ij}$ and $\sin\theta_{ij}$, respectively.
We obtain the following CKM matrix from the parameter set of Table~\ref{tb:yukawa}.
\begin{align}
    V_{CKM}=\begin{pmatrix}
    -0.841 - i 0.492 & 0.080 - i 0.210 &-0.0027 + i 0.0032\\
    -0.085 - i 0.208 &-0.827 + i 0.513 & 0.036 + i 0.019\\
    0.0039 - i 0.0054&-0.026 - i 0.032 & 0.335 - i 0.941\end{pmatrix}.
\end{align}
This is rewritten in accordance with the parametrization of Eq.~\eqref{eq:CKM} by a suitable rephasing operation as
\begin{align}
    V_{CKM}=\begin{pmatrix}
     0.974 & 0.225 & 0.00073 - i 0.00418\\
    -0.172 - i 0.144 & 0.746 + i 0.625 & 0.041\\
    0.0063 + i 0.0018&-0.031 - i 0.027 & 0.999\end{pmatrix}.
\end{align}
It leads to $\delta_{CP}=80.1^\circ$.
The absolute value of CKM is
\begin{align}
    |V_{CKM}|=\begin{pmatrix}
    0.974 & 0.225 & 0.0042\\
    0.224 & 0.973 & 0.041\\
    0.0066 & 0.041 & 0.999\end{pmatrix}.
\end{align}
We can see that the predicted CKM matrix elements are similar to its experimental data of at low energy scale.

It is useful to discuss CP violation with the rephasing invariant CP violating measure, the Jarlskog invariant~\cite{Jarlskog:1985ht}.
It is defined by the CKM matrix elements $V_{\alpha \beta}$. 
It is also written in terms of the mixing angles and the CP violating Dirac phase as
\begin{align}
    J_{CP}=\text{Im}\left [V_{ud}V_{cs}V_{us}^\ast U_{cd}^\ast \right ]
    =s_{23}c_{23}s_{12}c_{12}s_{13}c_{13}^2\sin \delta _\text{CP}~ .
\label{Jcp}
\end{align}
It is predicted to be $J_{CP}=2.445$, which is also consistent with the global fit of \cite{Ross:2007az} for $\tan{\beta}=50$.
%------------------------------------------------------------------------------%
%------------------------------------------------------------------------------%
%------------------------------------------------------------------------------%
\section{Summary}

We have investigated a model with $A_4$ flavor symmetry with $Z_3$, $U(1)_{FN}$ and $U(1)_R$ symmetries in the SUSY SU(5) framework.
It is required to modify $A_4$ flavor models in light of the latest experimental data of both quark and lepton mixings.
In our model, the 24-dimensional adjoint Higgs multiplet, $H_{24}$, makes a difference between masses of down-quarks and charged leptons.
Moreover, a FN field, $\Theta$, realizes the observed fermion mass hierarchy.
The flavons, $\phi_T$, $\phi_S$, $\xi$ and $\xi'$, are also introduced in order to realize the required neutrino oscillation parameters.
Since $\xi$ and $\xi'$ are coupled to the quarks not only to the leptons, these VEVs contribute to the CKM parameters too.
We have analyzed the VEVs of those scalar fields by the potential analysis and confirmed that we can obtain the required VEV alignment for flavor mixing.
In the numerical calculation, we have fixed the VEVs of $\xi,\xi'$ and $\Theta$ by hand.

We have shown the numerical result of quark mixing and mass ratios of quarks and charged leptons at GUT scale ($\sim10^{16}$ GeV).
Our result is consistent with the global fit data of running mass ratios among the quarks as well as charged leptons.
We have also shown the predicted CKM parameters and confirmed that the predicted CKM matrix is similar to its experimental data of at low energy scale.
The Jarlskog invariant $J_{CP}$ is also consistent with the data at GUT scale for $\tan{\beta}=50$.

%-------- acknowledgement -------%
\vspace{0.5cm}
\noindent

{\bf Acknowledgement}

Y.M. is supported in part by the National Natural Science Foundation of China (NNSFC) under Contracts Nos.~11675061, 11775092, 11521064 and 11435003.
Y.S. is supported by JSPS KAKENHI Grant Number JP17K05418 and Fujyukai Foundation.

%Thanks.

%---------------------------------------------------------%
%--------------- Appendix --------------------------------%
%---------------------------------------------------------%
\appendix
\section*{Appendix}
\section{Multiplication rule of $A_4$ group}
\label{sec:multiplication-rule}
We use the multiplication rule of the $A_4$ triplet as follows:
\begin{align}
\begin{pmatrix}
a_1\\
a_2\\
a_3
\end{pmatrix}_{\bf 3}
\otimes 
\begin{pmatrix}
b_1\\
b_2\\
b_3
\end{pmatrix}_{\bf 3}
&=\left (a_1b_1+a_2b_3+a_3b_2\right )_{\bf 1}
\oplus \left (a_3b_3+a_1b_2+a_2b_1\right )_{{\bf 1}'} \nonumber \\
& \oplus \left (a_2b_2+a_1b_3+a_3b_1\right )_{{\bf 1}''} \nonumber \\
&\oplus \frac13
\begin{pmatrix}
2a_1b_1-a_2b_3-a_3b_2 \\
2a_3b_3-a_1b_2-a_2b_1 \\
2a_2b_2-a_1b_3-a_3b_1
\end{pmatrix}_{{\bf 3}}
\oplus \frac12
\begin{pmatrix}
a_2b_3-a_3b_2 \\
a_1b_2-a_2b_1 \\
a_3b_1-a_1b_3
\end{pmatrix}_{{\bf 3}\  } \ , \nonumber \\
\nonumber \\
{\bf 1} \otimes {\bf 1} = {\bf 1} \ , \qquad &
{\bf 1'} \otimes {\bf 1'} = {\bf 1''} \ , \qquad
{\bf 1''} \otimes {\bf 1''} = {\bf 1'} \ , \qquad
{\bf 1'} \otimes {\bf 1''} = {\bf 1} \  .
\end{align}
More details are shown in the review~\cite{Ishimori:2010au,Ishimori:2012zz}.
%------------------------------------------------------------------------------%
%------------------------------------------------------------------------------%
%------------------------------------------------------------------------------%
\section{Model parameters of down-type}
We can reduce the degree of freedom of the model parameters by inputting the observed values of fermion mass ratios at the GUT scale.
We show  some relations among model parameters of the mass matrix in Eq.~\eqref{eq:Md} and the mass ratios for down-type quarks as well as that for the charged leptons.
At first, we introduce new parameters $k_i$ and parametrize the following Yukawa couplings as:
\begin{align}
	y_i^{\bar{5}}=y_i e^{i \phi_i^{\bar{5}}},\quad y_i\equiv |y_i^{\bar{5}}|,\quad
	y_i^{24}=k_iy_i e^{i \phi_i^{24}},\quad k_i\equiv |y_i^{24}|/|y_i^{\bar{5}}|.
\end{align}
By use of the parameters, the ratios of eigenvalues of the mass matrix in Eq.~\eqref{eq:Md} are written in the following forms:
\begin{align}
	\left(\frac{m_1}{m_2\lambda}\right)^2=\frac{y_1^2}{y_2^2}\frac{|1+\alpha k_1 e^{i\phi_1}|^2}{|1+\alpha k_2 e^{i\phi_2}|^2},\quad
	\left(\frac{m_2}{m_3\lambda^2}\right)^2=\frac{y_2^2}{y_3^2}\frac{|1+\alpha k_2 e^{i\phi_2}|^2}{|1+\alpha k_3 e^{i\phi_3}|^2},
\end{align}
where the lower indices denote the flavors and $\phi_i\equiv\phi_i^{24}-\phi_i^{\bar{5}}$.
They can be rewritten for the down-type quarks and charged leptons as
\begin{align}
	(\alpha_{d}k_2)^2+2\alpha_{d}k_2\cos{\phi_2}+1&=
	\left(\frac{m_s}{m_b\lambda^2}\right)^2\frac{y_3^2}{y_2^2}
	\left\{(\alpha_{d}k_3)^2+2\alpha_{d}k_3\cos{\phi_3}+1\right\},
	\label{eq:Bsb} \\
	(\alpha_{d}k_2)^2+2\frac{\alpha_d^2}{\alpha_c}k_2\cos{\phi_2}+\frac{\alpha_d^2}{\alpha_c^2}&=
	\frac{\alpha_d^2}{\alpha_c^2}\left(\frac{m_\mu}{m_\tau\lambda^2}\right)^2\frac{y_3^2}{y_2^2}
	\left\{(\alpha_{c}k_3)^2+2\alpha_{c}k_3\cos{\phi_3}+1\right\},
	\label{eq:Bmutau} \\
	(\alpha_{d}k_1)^2+2\alpha_{d}k_1\cos{\phi_1}+1&=
	\left(\frac{m_d}{m_s\lambda}\right)^2\frac{y_2^2}{y_1^2}
	\left\{(\alpha_{d}k_2)^2+2\alpha_{d}k_2\cos{\phi_2}+1\right\},
	\label{eq:Bds} \\
	(\alpha_{d}k_1)^2+2\frac{\alpha_d^2}{\alpha_c}k_1\cos{\phi_1}+\frac{\alpha_d^2}{\alpha_c^2}&=
	\frac{\alpha_d^2}{\alpha_c^2}\left(\frac{m_e}{m_\mu\lambda}\right)^2\frac{y_2^2}{y_1^2}
	\left\{(\alpha_{c}k_2)^2+2\alpha_{c}k_2\cos{\phi_2}+1\right\},
	\label{eq:Bemu}
\end{align}
where $\alpha_d$ and $\alpha_c$ denote the explicit value of $\alpha$ in Eq.~\eqref{eq:alpha} for down-type quarks and charged leptons respectively.
By subtracting the Eqs.~\eqref{eq:Bsb} and \eqref{eq:Bmutau}, we can write $k_2$ in terms of the specific mass ratios and some free parameters, $k_3$ and $y_i^{\bar 5}$ as:
\begin{align}
\begin{aligned}
	k_2=
%	&\frac1{2\alpha_d\cos{\phi_2}\left(1-\frac{\alpha_d}{\alpha_c}\right)}
%	\left[\frac{\alpha_d^2}{\alpha_c^2}-1+\frac{y_3^2}{y_2^2}
%	\left(\left(\frac{m_s}{m_b\lambda^2}\right)^2\left\{(\alpha_{d}k_3)^2+2\alpha_{d}k_3\cos{\phi_3}+1\right\}\right.\right. \\
%	&\left.\left.-\frac{\alpha_d^2}{\alpha_c^2}\left(\frac{m_\mu}{m_\tau\lambda^2}\right)^2
%	\left\{(\alpha_{c}k_3)^2+2\alpha_{c}k_3\cos{\phi_3}+1\right\}\right)\right]\\=
	&\frac1{5\sqrt{30}\cos{\phi_2}}\left[-\frac{25}2+\frac{y_3^2}{y_2^2\lambda^4}\left\{
	3\left(\left(\frac{m_s}{m_b}\right)^2-\left(\frac{m_\mu}{m_\tau}\right)^2\right)k_3^2 \right.\right. \\
	&+\left.\left.\sqrt{30}\left(3\left(\frac{m_s}{m_b}\right)^2
	+2\left(\frac{m_\mu}{m_\tau}\right)^2\right)k_3\cos{\phi_3}
	+\left(\frac{45}2\left(\frac{m_s}{m_b}\right)^2-10\left(\frac{m_\mu}{m_\tau}\right)^2\right)\right\}\right],
\end{aligned}
\end{align}
where the explicit values of $\alpha_d$ and $\alpha_c$ are substituted.
In the same manner, we can write $k_1$ in terms of the specific mass ratios, $k_2$ and $y_i^{\bar 5}$:
\begin{align}
\begin{aligned}
	k_1=
	&\frac1{5\sqrt{30}\cos{\phi_1}}\left[-\frac{25}2+\frac{y_2^2}{y_1^2\lambda^2}\left\{
	3\left(\left(\frac{m_d}{m_s}\right)^2-\left(\frac{m_e}{m_\mu}\right)^2\right)k_2^2 \right.\right. \\
	&+\left.\left.\sqrt{30}\left(3\left(\frac{m_d}{m_s}\right)^2
	+2\left(\frac{m_e}{m_\mu}\right)^2\right)k_2\cos{\phi_2}
	+\left(\frac{45}2\left(\frac{m_d}{m_s}\right)^2-10\left(\frac{m_e}{m_\mu}\right)^2\right)\right\}\right].
\end{aligned}
\end{align}
Therefore, we can obtain the magnitudes of $y_{i}^{24}$ automatically by inputting $y_{i}^{\bar 5}$ and $k_3$ with the observed values of fermion mass ratios.
%------------------------------------------------------------------------------%
%--------------------------    References    ----------------------------------%
%------------------------------------------------------------------------------%
%\newpage


\begin{thebibliography}{99}


%----------  Review  ----------%
%\cite{Altarelli:2010gt}
\bibitem{Altarelli:2010gt}
G.~Altarelli and F.~Feruglio,
%``Discrete Flavor Symmetries and Models of Neutrino Mixing,''
Rev.\ Mod.\ Phys.\  {\bf 82} (2010) 2701
%doi:10.1103/RevModPhys.82.2701
[arXiv:1002.0211 [hep-ph]].



%\cite{Ishimori:2010au}
\bibitem{Ishimori:2010au}
H.~Ishimori, T.~Kobayashi, H.~Ohki, Y.~Shimizu, H.~Okada and M.~Tanimoto,
%``Non-Abelian Discrete Symmetries in Particle Physics,''
Prog.\ Theor.\ Phys.\ Suppl.\  {\bf 183} (2010) 1
[arXiv:1003.3552 [hep-th]].
%%CITATION = ARXIV:1003.3552;%%



%\cite{Ishimori:2012zz}
\bibitem{Ishimori:2012zz}
H.~Ishimori, T.~Kobayashi, H.~Ohki, H.~Okada, Y.~Shimizu and M.~Tanimoto,
%``An introduction to non-Abelian discrete symmetries for particle physicists,''
Lect.\ Notes Phys.\  {\bf 858} (2012) 1, Springer.
%%CITATION = LNPHA,858,1;%%


%\cite{King:2013eh}
\bibitem{King:2013eh}
S.~F.~King and C.~Luhn,
%``Neutrino Mass and Mixing with Discrete Symmetry,''
Rept.\ Prog.\ Phys.\  {\bf 76} (2013) 056201
% doi:10.1088/0034-4885/76/5/056201
[arXiv:1301.1340 [hep-ph]].

%\cite{King:2014nza}
\bibitem{King:2014nza}
S.~F.~King, A.~Merle, S.~Morisi, Y.~Shimizu and M.~Tanimoto,
%``Neutrino Mass and Mixing: from Theory to Experiment,''
arXiv:1402.4271 [hep-ph].


%----------  A4 & SUSY SU(5) model  ----------%



%----------  CKM fitter  ----------%
%\cite{Charles:2004jd}
\bibitem{Charles:2004jd}
J.~Charles {\it et al.} [CKMfitter Group],
%``CP violation and the CKM matrix: Assessing the impact of the asymmetric $B$ factories,''
Eur.\ Phys.\ J.\ C {\bf 41} (2005) no.1,  1
%doi:10.1140/epjc/s2005-02169-1
[hep-ph/0406184].
%%CITATION = doi:10.1140/epjc/s2005-02169-1;%%
%1641 citations counted in INSPIRE as of 01 Oct 2018


%----------  Ross Serna  ----------%
%\cite{Ross:2007az}
\bibitem{Ross:2007az}
G.~Ross and M.~Serna,
%``Unification and fermion mass structure,''
Phys.\ Lett.\ B {\bf 664} (2008) 97
%doi:10.1016/j.physletb.2008.05.014
[arXiv:0704.1248 [hep-ph]].
%%CITATION = doi:10.1016/j.physletb.2008.05.014;%%
%117 citations counted in INSPIRE as of 01 Oct 2018

%\cite{Buras:1977yy}
\bibitem{Buras:1977yy}
A.~J.~Buras, J.~R.~Ellis, M.~K.~Gaillard and D.~V.~Nanopoulos,
%``Aspects of the Grand Unification of Strong, Weak and Electromagnetic Interactions,''
Nucl.\ Phys.\ B {\bf 135}, 66 (1978).
%doi:10.1016/0550-3213(78)90214-6
%%CITATION = doi:10.1016/0550-3213(78)90214-6;%%
%1313 citations counted in INSPIRE as of 25 May 2019

%\cite{Ellis:1979fg}
\bibitem{Ellis:1979fg}
J.~R.~Ellis and M.~K.~Gaillard,
%``Fermion Masses and Higgs Representations in SU(5),''
Phys.\ Lett.\  {\bf 88B}, 315 (1979).
%doi:10.1016/0370-2693(79)90476-3
%%CITATION = doi:10.1016/0370-2693(79)90476-3;%%
%231 citations counted in INSPIRE as of 23 May 2019



%----------  A4  modified  ----------%
%\cite{Shimizu:2011xg}
\bibitem{Shimizu:2011xg}
Y.~Shimizu, M.~Tanimoto and A.~Watanabe,
%``Breaking Tri-bimaximal Mixing and Large $\theta_{13}$,''
Prog.\ Theor.\ Phys.\  {\bf 126} (2011) 81
%doi:10.1143/PTP.126.81
[arXiv:1105.2929 [hep-ph]].
%%CITATION = doi:10.1143/PTP.126.81;%%


%----------  NuFIT   ----------%
%\cite{Esteban:2016qun}
\bibitem{Esteban:2016qun}
I.~Esteban, M.~C.~Gonzalez-Garcia, M.~Maltoni, I.~Martinez-Soler and T.~Schwetz,
%``Updated fit to three neutrino mixing: exploring the accelerator-reactor complementarity,''
JHEP {\bf 1701} (2017) 087
%doi:10.1007/JHEP01(2017)087
[arXiv:1611.01514 [hep-ph]],
NuFIT 3.2 (2018), www.nu-fit.org
%%CITATION = doi:10.1007/JHEP01(2017)087;%%
%370 citations counted in INSPIRE as of 01 Oct 2018

%\cite{Georgi:1974sy}
\bibitem{Georgi:1974sy}
H.~Georgi and S.~L.~Glashow,
%``Unity of All Elementary Particle Forces,''
Phys.\ Rev.\ Lett.\  {\bf 32}, 438 (1974).
%doi:10.1103/PhysRevLett.32.438

%%CITATION = doi:10.1103/PhysRevLett.32.438;%%
%4499 citations counted in INSPIRE as of 08 Feb 2018


%\cite{Antusch:2011qg}
\bibitem{Antusch:2011qg}
  S.~Antusch and V.~Maurer,
  %``Large neutrino mixing angle $\theta_{13}$^{MNS} and quark-lepton mass ratios in unified flavour models,''
  Phys.\ Rev.\ D {\bf 84} (2011) 117301
  %doi:10.1103/PhysRevD.84.117301
  [arXiv:1107.3728 [hep-ph]].
  %%CITATION = doi:10.1103/PhysRevD.84.117301;%%
  %76 citations counted in INSPIRE as of 03 Jun 2019
  
  

%\cite{King:2017guk}
\bibitem{King:2017guk}
S.~F.~King,
%``Unified Models of Neutrinos, Flavour and CP Violation,''
Prog.\ Part.\ Nucl.\ Phys.\  {\bf 94} (2017) 217
%doi:10.1016/j.ppnp.2017.01.003
[arXiv:1701.04413 [hep-ph]].
%%CITATION = doi:10.1016/j.ppnp.2017.01.003;%%
%62 citations counted in INSPIRE as of 31 May 2019



  
%----------   AFH   ----------%
%\cite{Altarelli:2008bg}
\bibitem{Altarelli:2008bg}
G.~Altarelli, F.~Feruglio and C.~Hagedorn,
%``A SUSY SU(5) Grand Unified Model of Tri-Bimaximal Mixing from A$_4$,''
JHEP {\bf 0803} (2008) 052
%doi:10.1088/1126-6708/2008/03/052
[arXiv:0802.0090 [hep-ph]].
%%CITATION = doi:10.1088/1126-6708/2008/03/052;%%
%174 citations counted in INSPIRE as of 01 Oct 2018


%----------  Reactor ----------%
%\cite{An:2012eh}
\bibitem{An:2012eh}
F.~P.~An {\it et al.}  [DAYA-BAY Collaboration],
%``Observation of electron-antineutrino disappearance at Daya Bay,''
Phys.\ Rev.\ Lett.\  {\bf 108} (2012) 171803
[arXiv:1203.1669 [hep-ex]].

%\cite{Ahn:2012nd}
\bibitem{Ahn:2012nd}
J.~K.~Ahn {\it et al.}  [RENO Collaboration],
%``Observation of Reactor Electron Antineutrino Disappearance in the RENO Experiment,''
Phys.\ Rev.\ Lett.\  {\bf 108} (2012) 191802  [arXiv:1204.0626 [hep-ex]].


%-------------------------%

%\cite{Bjorkeroth:2015ora}
\bibitem{Bjorkeroth:2015ora}
F.~Bj\"{o}rkeroth, F.~J.~de Anda, I.~de Medeiros Varzielas and S.~F.~King,
%``Towards a complete A$_{4} \times$ SU(5) SUSY GUT,''
JHEP {\bf 1506} (2015) 141
%doi:10.1007/JHEP06(2015)141
[arXiv:1503.03306 [hep-ph]].
%%CITATION = doi:10.1007/JHEP06(2015)141;%%
%54 citations counted in INSPIRE as of 31 May 2019

%\cite{Belyaev:2018vkl}
\bibitem{Belyaev:2018vkl}
  A.~S.~Belyaev, S.~F.~King and P.~B.~Schaefers,
  %``Muon g-2 and dark matter suggest nonuniversal gaugino masses: $\mathbf{SU(5)\times A_4}$ case study at the 
LHC,''
  Phys.\ Rev.\ D {\bf 97} (2018) no.11,  115002
  %doi:10.1103/PhysRevD.97.115002
  [arXiv:1801.00514 [hep-ph]].
  %%CITATION = doi:10.1103/PhysRevD.97.115002;%%
  %15 citations counted in INSPIRE as of 03 Jun 2019
  
\cite{Laamara:2018zpo}
\bibitem{Laamara:2018zpo}
  R.~A.~Laamara, M.~A.~Loualidi, M.~Miskaoui and E.~H.~Saidi,
  %``Hybrid seesaw neutrino model in SUSY $SU(5)\times \mathbb{A}_{4}$,''
  Phys.\ Rev.\ D {\bf 98} (2018) no.1,  015004
  %doi:10.1103/PhysRevD.98.015004
  [arXiv:1806.08573 [hep-ph]].
  %%CITATION = doi:10.1103/PhysRevD.98.015004;%%
  
%\cite{Bernigaud:2018qky}
\bibitem{Bernigaud:2018qky}
  J.~Bernigaud, B.~Herrmann, S.~F.~King and S.~J.~Rowley,
  %``Non-minimal flavour violation in A$_{4} \times$ SU(5) SUSY GUTs with smuon assisted dark matter,''
  JHEP {\bf 1903} (2019) 067
  %doi:10.1007/JHEP03(2019)067
  [arXiv:1812.07463 [hep-ph]].
  %%CITATION = doi:10.1007/JHEP03(2019)067;%%
  %2 citations counted in INSPIRE as of 03 Jun 2019
  
  

%\cite{Froggatt:1978nt}
\bibitem{Froggatt:1978nt}
C.~D.~Froggatt and H.~B.~Nielsen,
%``Hierarchy of Quark Masses, Cabibbo Angles and CP Violation,''
Nucl.\ Phys.\ B {\bf 147} (1979) 277.
%doi:10.1016/0550-3213(79)90316-X
%%CITATION = doi:10.1016/0550-3213(79)90316-X;%%
%1646 citations counted in INSPIRE as of 31 May 2019
%-----------@ seesaw  --------------%

%\cite{Minkowski:1977sc}
\bibitem{Minkowski:1977sc} 
P.~Minkowski,
%``$\mu \to e\gamma$ at a Rate of One Out of $10^{9}$ Muon Decays?,''
Phys.\ Lett.\  {\bf 67B}, 421 (1977); 
%doi:10.1016/0370-2693(77)90435-X
%%CITATION = doi:10.1016/0370-2693(77)90435-X;%%
%3395 citations counted in INSPIRE as of 23 May 2019


%\cite{Yanagida}
\bibitem{Yanagida} 
T. Yanagida, in Proceedings of the Workshop on Unified Theories and Baryon Number in the Universe, eds. O. Sawada and A. Sugamoto (KEK report 79-18, 1979); 


%\cite{GellMann:1980vs}
\bibitem{GellMann:1980vs} 
M.~Gell-Mann, P.~Ramond and R.~Slansky,
%``Complex Spinors and Unified Theories,''
Conf.\ Proc.\ C {\bf 790927}, 315 (1979)
[arXiv:1306.4669 [hep-th]]; 
%%CITATION = ARXIV:1306.4669;%%
%2871 citations counted in INSPIRE as of 23 May 2019


%\cite{Mohapatra:1979ia}
\bibitem{Mohapatra:1979ia} 
R.~N.~Mohapatra and G.~Senjanovic,
%``Neutrino Mass and Spontaneous Parity Nonconservation,''
Phys.\ Rev.\ Lett.\  {\bf 44}, 912 (1980); 
%doi:10.1103/PhysRevLett.44.912
%%CITATION = doi:10.1103/PhysRevLett.44.912;%%
%4961 citations counted in INSPIRE as of 23 May 2019


%\cite{Schechter:1980gr}
\bibitem{Schechter:1980gr} 
J.~Schechter and J.~W.~F.~Valle,
%``Neutrino Masses in SU(2) x U(1) Theories,''
Phys.\ Rev.\ D {\bf 22}, 2227 (1980).
%doi:10.1103/PhysRevD.22.2227
%%CITATION = doi:10.1103/PhysRevD.22.2227;%%
%2513 citations counted in INSPIRE as of 23 May 2019


%\cite{Morozumi:2017rrg}
\bibitem{Morozumi:2017rrg}
  T.~Morozumi, H.~Okane, H.~Sakamoto, Y.~Shimizu, K.~Takagi and H.~Umeeda,
  %``Phenomenological Aspects of Possible Vacua of a Neutrino Flavor Model,''
  Chin.\ Phys.\ C {\bf 42} (2018) no.2,  023102
 %doi:10.1088/1674-1137/42/2/023102
  [arXiv:1707.04028 [hep-ph]].
  %%CITATION = doi:10.1088/1674-1137/42/2/023102;%%
  %5 citations counted in INSPIRE as of 31 May 2019


%%%%%%%%%%%%%%%%%%%%%%%%%%%%%%%%%%%%%%%%%%%%%%%%%%%
%%\cite{Haba:1999fk}
%\bibitem{Haba:1999fk}
%N.~Haba and N.~Okamura,
%%``Stability of the lepton-flavor mixing matrix against quantum corrections,''
%Eur.\ Phys.\ J.\ C {\bf 14} (2000) 347
%%doi:10.1007/s100520000333
%[hep-ph/9906481].


%\cite{Goto:1998qg}
\bibitem{Goto:1998qg} 
T.~Goto and T.~Nihei,
%``Effect of RRRR dimension five operator on the proton decay in the minimal SU(5) SUGRA GUT model,''
Phys.\ Rev.\ D {\bf 59}, 115009 (1999)
%doi:10.1103/PhysRevD.59.115009 [hep-ph/9808255].
%%CITATION = doi:10.1103/PhysRevD.59.115009;%%
%238 citations counted in INSPIRE as of 28 May 2019




%----------  JCP  ----------%

%\cite{Jarlskog:1985ht}
\bibitem{Jarlskog:1985ht}
C.~Jarlskog,
%``Commutator of the Quark Mass Matrices in the Standard Electroweak Model and a Measure of Maximal CP Violation,''
Phys.\ Rev.\ Lett.\  {\bf 55} (1985) 1039.
%%CITATION = PRLTA,55,1039;%%
%1141 citations counted in INSPIRE as of 22 Apr 2014


\end{thebibliography}
\end{document}